\documentclass[12pt,a4paper]{article}

\newcommand\be{\begin{equation}}
\newcommand\ee{\end{equation}}
\newcommand\bea{\begin{eqnarray}}
\newcommand\eea{\end{eqnarray}}
\newcommand\ket[1]{|#1\rangle}
\newcommand\bra[1]{\langle #1|}

\newcommand{\fatalpha}{{\bf \alpha \kern -0.44em \alpha}}
\newcommand{\fatsigma}{{\bf \sigma \kern -0.54em \sigma}}
\newcommand{\tpchi}{{\bf \chi \kern -0.35em \chi}}
\newcommand{\llambda}{{\bf \lambda \kern -0.45em \lambda}}

\newcommand \clebsch[6]{\left<\begin{array}{cc|c}
   #1 & #2 & #3 \\ #4 & #5 & #6 \end{array} \right>} 

\begin{document}
\title{Contracting the Wigner-Kernel of a Spin to the Wigner-Kernel of a Particle}
\author{Jean-Pierre Amiet and Stefan Weigert \\
Institut de Physique, Universit\'e de Neuch\^atel\\ 
Rue A.-L.\ Breguet 1, CH-2000 Neuch\^atel, Switzerland\\ 
\tt stefan.weigert@iph.unine.ch}
\date{April 2000}
\maketitle

\begin{abstract}
A general relation between the Moyal formalisms for a spin and a particle 
is established. Once the formalism has been set up for a spin, the phase-space description of a particle is obtained from the `contraction' of the group of rotations to the group of translations. This is shown by explicitly contracting a spin Wigner-kernel to the Wigner kernel of a particle. In fact, only {\em one} out of $2^{2s}$ different possible kernels for a spin shows this behaviour.
\end{abstract}

\section{Introduction}

To represent quantum mechanics in terms of $c$-number valued 
functions has various appealing properties. It 
becomes possible to situate the quantum mechanical description 
of a system in a familiar frame, namely the phase space of its
classical analog. Similarities and differences of the two descriptions 
can be visualized particularly well in such an approach. Further, from a structural point of view, to calculate expectation values of operators 
by means of `quasi-probabilities' in phase space, is  strongly analogous to the determination of mean values in classical statistical mechanics \cite{wigner32}. 
The basic ingredient to set up such a {\em symbolic calculus} is a one-to-one correspondence between (self-adjoint) operators $\widehat A$ (acting on a Hilbert space $\cal H$) on the one hand, and (real) functions $W_A$ defined on the phase-space $\Gamma$ of the classical system on the other. 

The quantum mechanics of spin and particle systems can be
represented faithfully in terms of functions defined on the surface 
of a sphere with radius $s$, and on a plane, respectively. Intuitively, one 
expects these
phase space-formulations to approach each other for increasing values of the
spin quantum number since the surface of a sphere is then
approximated by a plane with increasing accuracy. Therefore, appropriate
Wigner functions of a spin, say, should go over smoothly into
particle Wigner-functions in the limit of large $s$. It will be shown
how this transition can be transformed in a rigorous and
general way. The derivation is based on the group theoretical
technique of {\em contraction}. The group $SU(2)$ of quantum mechanical rotations is contracted to the Heisenberg-Weyl group $HW_1$ associated with the particle. In this procedure, rotations go over into translations. Subsequently, the operator kernel which defines the spin Wigner-formalism in a condensed manner will be shown to contract to the operator kernel for a particle in the limit of infinite $s$.
\section{Wigner-kernel for a particle}

Consider a particle on the real line $I \!\! R^1$, with position and momentum operators satisfying $[\hat q , \hat p] = i\hbar$.  
The Stratonovich-Weyl correspondence, associating operators with functions in phase space, can be characterized elegantly by means of a {\em kernel} \cite{royer77,huguenin+81},
\be \label{particlekernel}
 \widehat{ \Delta}(\alpha) = 2 \widehat{T} (\alpha) \, {\widehat \Pi }
  \, \widehat{T}^\dagger (\alpha) \, ,
                    \qquad \alpha =\frac{1}{\sqrt{2}}(q+ip) \in \Gamma \equiv C\!\!\!\!l \, ,
\ee
which has an interpretation as a {\em parity operator} displaced by $\alpha$. The unitary \cite{perelomov86}
\be \label{trlop}
\widehat{ T} (\alpha) =  \exp [ \alpha a^+ - \alpha^* a ] \, , 
\ee
effects translations in phase space $\Gamma$, 
\be \label{trlopeffect}
a \rightarrow \widehat{ T} (\alpha) a \widehat{ T}^\dagger (\alpha) 
       = a - \alpha \, , 
\ee
where $a^-\equiv a = (\hat q - i \hat p)/\sqrt{2} $ and $a^+ = a^\dagger$ are the standard annihilation and creation operators ($\hbar =1$).
At the origin $\alpha = 0$, the kernel equals (two times) the unitary, involutive parity operator $\widehat \Pi$, 
\be \label{parityeffect}
{\widehat \Pi } \, a  \, {\widehat \Pi }^\dagger = - a \, ,
\ee
corresponding to a reflection at the origin of $\Gamma$. Using the number operator $\widehat N =a^+a$ and its eigenstates, 
\be \label{numbereigen}
\widehat N \ket{n} = n \ket{n} \, ,  \qquad n = 0,1,2, \ldots \, ,
\ee
parity can be given a simple form which will be useful later,
\be \label{parityforms}
{\widehat \Pi } 
      = \exp [ i\pi \widehat N ] 
      = \sum_{n=0}^\infty (-)^n \ket{n}\bra{n} \, .
\ee
The kernel $\widehat{ \Delta}(\alpha)$ can be derived from the {\em Stratonovich-Weyl} postulates \cite{stratonovich57} which are natural conditions  on a quantum mechanical phase-space representation.
The correspondence between a (self-adjoint) operator $\widehat A$  and a (real) function is defined by
\be \label{newwigner}
W_{A} (\alpha) 
       = \mbox{ Tr }  \left[ \widehat{ \Delta}(\alpha) \widehat A \right] \, ,
\ee
while its inverse reads
\be  \label{newwignerinverse}
\widehat A = \int_\Gamma d\alpha \, 
                  W_{A} (\alpha) \widehat{ \Delta}(\alpha) \, .
\ee
If $\widehat A$ is the density operator  of a pure state, 
$\hat \rho = \ket{\psi}\bra{\psi}$, the symbol defined in (\ref{newwigner}) 
is the {\em Wigner function} of the state $\ket{\psi}$,
\be \label{oldwigner}
W_{\psi}(p,q) = \frac{2}{h}\int_\Gamma \, dx \, \psi^* (q+x) \psi
                                                 (q-x) \exp[2ipx/\hbar] \, .
\ee
It is important to note that the kernel $\widehat \Delta (\alpha)$ is entirely defined in terms of the operators $a^\pm$ and $\widehat N$, forming a closed algebra under commutation if the identity is included:
\be \label{HWalg}
[ a , a^+ ] = 1\, , \qquad 
[\widehat N , a^\pm ] = \pm a^\pm \, .
\ee
This algebra generates the {\em Heisenberg-Weyl group} $HW_1$, and the kernel
$\widehat{ \Delta}(\alpha)$ is an element of it (apart from the factor of two).
\section{Wigner-kernel for a spin}
For a quantum spin, the symbol associated with an operator is a
continuous function defined on the {\em sphere} ${\cal S}^2$, being the
phase space of the classical spin. When setting up a phase-space formalism,  rotations take over the role of translations.  The group $SU(2)$ is generated by the components of the spin operator $\widehat {\bf S}$. The three operators $\widehat S^\pm = (\widehat S^x \pm i \widehat S^y) $ and $\widehat S^z$, satisfy the commutation relations 
\be \label{su2}
[ \widehat S^+ , \widehat S^- ] = 2 \widehat S^z \, , \qquad 
[\widehat S^z ,\widehat S^\pm] = \widehat S^\pm \, .
\ee
The standard basis 
\be \label{zbasis}
{\bf n}_z  \cdot  {\bf \hat{S}}  \ket{s,m} = m \ket{s,m} \, , \qquad 
                                                             m = -s, \ldots , s \, ,
\ee
is given by the eigenstates of the $z$ component $ \widehat S^z$ of the spin.

For a quantum spin, it is natural to expect that the elements of the 
Wigner kernel will be labeled by points of the sphere ${\cal S}^2$, corresponding to unit vectors ${\bf n}$ $= (\sin \vartheta \cos \varphi,$$\sin \vartheta \sin \varphi , \cos \vartheta )$, parametrized by standard spherical coordinates. Replacing intuitively translations in (\ref{particlekernel}) by rotations leads to the expression 
\be \label{suchaniceform}
\widehat\Delta({\bf n})
            = \widehat U({\bf n})  \, 
             \widehat \Pi_{s} \, \widehat U^\dagger ({\bf n}) \, ,
\ee
where 
\be \label{rotation}
\widehat U ({\bf n}) 
= \exp [- i \vartheta {\bf k} \cdot \widehat {\bf S} ] 
\ee
with a unit vector ${\bf k} = (-\sin \varphi, \cos \varphi, 0 )$ in the $xy$ plane. Thus, $\widehat U ({\bf n})$ represents a finite rotation which maps the operator $\widehat S^z = {\bf n}_z \cdot \widehat {\bf S}$ into ${\bf n} \cdot \widehat {\bf S}$, i.e. ${\bf n}_z  \rightarrow {\bf n}$. What are natural choices for the operator  $\widehat \Pi_{s}$? 

Two possibilities come to one's mind. First, try to transfer the concept of reflection about some point in phase space. Introduce canonical coordinates $(q,p) = ( \varphi, \cos \vartheta)$ on the sphere. Then, `parity'  would  correspond to the map $(\varphi, \cos \vartheta) \to ( -\varphi, - \cos \vartheta)$, or $(\varphi, \vartheta) \to (2\pi - \varphi, \pi - \vartheta)$. This is just a rotation by $\pi$ about the $x$ axis. Since all points of the sphere are equivalent, one could also chose a rotation by $\pi$ about the $z$ axis as candidate for parity. Second, $\widehat \Pi_{s}$ might be considered to generate reflections about the center of the sphere, ${\bf n} \rightarrow -{\bf n} $, that is, $(\varphi, \vartheta) \rightarrow (\varphi + \pi, \pi - \vartheta)$. It can be shown that {\em both} possibilities do {\em not} give rise to a symbolic calculus on the sphere \cite{hilti98}, violating bijectivity between operators and phase-space functions, for example.

Nevetheless, acceptable operator kernels $\widehat\Delta_\varepsilon ({\bf n}) $ {\em do} exist as shown by Stratonovich \cite{stratonovich57}, V\'arilly and Gracia-Bond\'\i a \cite{varilly+89}, and by Amiet and Cibils \cite{amiet+91}. For example, the 
condition that the kernel should satisfy appropriate Stratonovich-Weyl postulates implies \cite{varilly+89} that 

\be \label{varillyansatz}
{\widehat \Delta_\varepsilon  ({\bf n}) }
     = \sum_{m,m'=-s}^s Z_{mm'}^\varepsilon ({\bf n}) \ket{s, m} \bra{s, m'} \, .
\ee
The coefficients, 
\be \label{eqSolBrute}
  Z_{mm^\prime}^\varepsilon ({\bf n})
         = \frac{\sqrt{4\pi}}{2s+1} \sum_{l=0}^{2s} \varepsilon_l \sqrt{2l+1}
             \clebsch{s}{l}{s}{m}{m^\prime-m}{m^\prime}
              Y_{l, m^\prime-m}({\bf n}) \, ,
\ee
where  $ \varepsilon_0=1 $ and $ \varepsilon_l=\pm 1 \, , l=1,
\ldots, 2s$, are linear combinations of Clebsch-Gordan coefficients multiplied by spherical harmonics  $Y_{l, m}({\bf n}), l= 0, 1, \ldots , 2s,$ $ m = -l, \ldots , l$. Note that there is {\em no} unique kernel but, due to the factors 
$ \varepsilon_l$, one can define $2^{2s} $ different Stratonovich-Weyl correspondence rules. 

Unfortunatley, the expression (\ref{varillyansatz}) does not admit a simple interpretation of the operator in analogy to (\ref{particlekernel}). It follows from  an independent derivation \cite{heiss+00/1} of $\widehat\Delta({\bf n}) $ that (\ref{varillyansatz}) can be written in the form (\ref{suchaniceform}) where 
\be \label{suchaniceform2}
 \widehat \Pi_s = \widehat\Delta_\varepsilon ({\bf n}_z)
                 = \sum_{m=-s}^s \Delta_\varepsilon (m) \ket{s, m} \bra{s, m} \, , 
\ee
with coefficients 
\be \label{simplecoeff}
\Delta_\varepsilon (m) = \sum_{l=0}^{2s} \varepsilon_l \frac{2l+1}{2s+1}
                                           \clebsch{s}{l}{s}{m}{0}{m} \, .
\ee
Still, the operator $\widehat \Pi_s $ does not have an obvious interpretation but a new strategy to justify its form emerges. Consider a plane tangent to the sphere at its north pole. For increasing radius, the sphere is approximated locally better and better by the plane. Therefore, one might expect that for $s \rightarrow \infty$ objects defined on the sphere turn into objects defined on the plane. It has been conjectured in \cite{heiss+00/1} that in this limit the Wigner kernel of a spin goes over into the kernel for a particle. It is the purpose of this paper to show that
\be \label{contract!}
\lim_{s \to \infty} \widehat U({\bf n}) \,  
             \widehat \Delta ({\bf n}_z) \,  \widehat U^\dagger ({\bf n})
                             =  \widehat \Delta (\alpha)  \, ,
 \ee
is indeed true for the kernel $\widehat \Delta_\varepsilon ({\bf n}_z)$ with parameters $\varepsilon_1= \varepsilon_2= \ldots =\varepsilon_{2s}=1$, denoted by $\widehat \Delta ({\bf n}_z)$ for short. Thus, while the rotations $\widehat U ({\bf n})$ should go over into translations, the operator $\widehat\Delta({\bf n}_z) $ corresponds, in one way or another, to parity for a spin. A convenient framewok to prove (\ref{contract!}) is the {\em contraction} of groups \cite{arecchi+72} as shown in the next section.  
\section{Contracting $SU(2)$ to $HW_1$} 
Introduce three operators $\widehat A^\pm$ and $\widehat A^z$  defined as linear combinations of the generators of the algebra $su(2)$ in polar form, 
\be \label{contrtrf}
\widehat A^\pm = c \widehat S^\mp \, , \qquad 
\widehat A^z =  -\widehat S^z + \frac{1_s}{2c^2}  \, , 
\ee
plus the identity $1_s$. This transformation is invertible for each value of the parameter $c > 0$. The non-zero commutators of the new generators are given by 
\be \label{newsu2}
[ \widehat A^- , \widehat A^+ ] = 1_s  - 2 c^2\widehat A^z \, , \qquad 
[\widehat A^\pm ,\widehat A^z] = \widehat A^\pm \, .
\ee
These relations have a well defined limit if $c \to 0$, nonwithstanding that 
the transformation (\ref{contrtrf}) is not invertible for $c=0$. In fact, they reproduce the commutation relations (\ref{HWalg}) of the Heisenberg-Weyl algebra after identifying
\be \label{identify}
\lim_{c\to 0} \widehat A^\pm = a^\pm \, , \qquad 
 \lim_{c\to 0} \widehat A^z     = \widehat N \, , \qquad
\lim_{c\to 0} 1_s = 1\, .
\ee
How do rotations behave in this limit? Any finite rotation 
$\widehat U({\bf n})\in SU(2) $ in (\ref{rotation}) can be written in the form 
\be \label{finiteu2}
\widehat U ({\bf n}) 
= \exp \left[ \xi_-\widehat S^ - - \xi_+ \widehat S^+ \right] \, , \qquad 
\xi_- = \frac{\vartheta}{2} e^{i\varphi} \, , \quad \xi_+ = \xi_-^* \, ,
\ee
or, expressed in terms of the operators (\ref{contrtrf}),
\be \label{finiteu3}
\widehat U ({\bf n}) 
= \exp \left[ c ( \xi_- \widehat A^+ -  \xi_+ \widehat A^-)\right ] \, .
\ee
Consequently, if  the coefficients $\xi_\pm$ shrink with the parameter $c$ according to  
\be \label{lambdalimit}
\lim_{c \to 0} \frac{\xi_-}{c}   
          = \lim_{c \to 0} \frac{\vartheta e^{i\varphi}}{2c} 
          =  \alpha \, , \qquad 
\lim_{c \to 0} \frac{\xi_+}{c}   
          = \lim_{c \to 0} \frac{\vartheta e^{-i\varphi}}{2c} 
          =  \alpha^* \, , 
\ee
a rotation  $\widehat U ({\bf n}) $ tends to a well-defined 
element of the Heisenberg-Weyl group,  Eq.\ (\ref{trlop}): 
\be \label{limitrot}
\lim_{c \to 0}  \widehat U ({\bf n})  = \widehat T (\alpha) \, .
\ee 
For consistency, the limit $c \to 0$ must correctly reproduce the eigenvalues of the operator $\widehat N$, given by the non-negative integers. Let us look at the fate of the eigenvalue equation (\ref{zbasis}) for $m= s$, which is expected to give $\widehat N \ket{n} =0$. One has 
\be \label{zbasisfate}
\lim_{c\to 0} \left(\widehat A^z \ket{s,s} \right)  
\lim_{c\to 0} \left[\left(- \widehat S^z + \frac{1_s}{2c^2} \right) \ket{s,s} \right]
     = \lim_{c\to 0} \left(- s + \frac{1}{2c^2} \right) \lim_{c\to 0} \ket{s,s} = 0 \,  
\ee
implying $2 c^2  s = 1$ for $\lim_{c\to 0} \ket{s,s} = \ket{n=0}$. Consequently, the radius of the sphere, $s$, increases with decreasing values of $c$. The state $\ket{s,s}$ turns indeed into the ground state associated with the operator $\widehat N$ since one has in general  
\be \label{limiteigenstates}
\lim_{c \to 0} \ket{s,m} = \lim_{c \to 0} \ket{s,s-n} =  \ket{n} \, ,  
\qquad n = s-m \in I \!\!\!{N_0} \, ,
\ee
as follows from 
\be \label{test}
\widehat N \ket{n} 
     = \lim_{c \to 0}  \left[ 
                         \left( - \widehat S^z + \frac{1_s}{2c^2} \right) \ket{s,s-n}\right]
     = \lim_{c \to 0} \left(  (s- m) + (\frac{1}{2c^2}  - s)\right) \ket{n}
     = n \ket{n} \, .
\ee
Now it is obvious why one needs  to associate the {\em creation} operator $\widehat S^+$ with the {\em annihilation} operator $a$ (cf. (\ref{contrtrf})): the eigenstates with {\em maximal} $s$ are linked to the oscillator ground state with {\em minimal} $n=0$. In \cite{arecchi+72}, a different convention has been used. Nevertheless,
it remains true that not only spin eigenstates are mapped into number eigenstates but many other expressions related to the group $U(2)$ turn into 
an equivalent expression for the group $HW_1$. 

This is good news for the present purpose to establish a relation between the Moyal formalism of a particle and a spin. Consider the limit of the kernel (\ref{suchaniceform}) under contraction using (\ref{limitrot}), 
\be \label{suchaniceform3}
\lim_{c\to 0} \widehat\Delta ({\bf n})
            = \widehat T(\alpha)  
                  \left( \lim_{c \to 0} \widehat \Pi_s\right) 
                \widehat T^\dagger (\alpha) \, .
\ee
The middle term can be written as
\be \label{finite?}
\lim_{c \to 0} \widehat \Pi_s
       = \lim_{c \to 0} \, \, \sum_{m=-s}^s \Delta_\varepsilon (m) \ket{s, m} \bra{s, m}
       = \sum_{n=0}^\infty 
              \left( \lim_{c \to 0} \Delta_\varepsilon (s-n) \right) \ket{n} \bra{n} \, .
\ee
Upon comparison with (\ref{parityforms}), the Wigner kernel of a spin is seen to turn into the Wigner kernel of the particle if 
\be \label{sumtoprove}
\lim_{s\to \infty} \, \, \sum_{l=0}^{2s} \varepsilon_l
                 \left(\frac{2l+1}{2s+1}\right)^{1/2} 
                 \clebsch{s}{s}{l}{s-n}{n-s}{0} = 2 
\ee
holds for all non-negative integers $n$. In the next section, this will be shown to be true for the choice $\varepsilon_l =+1$, $l=1, \ldots 2s$.
 \section{Summing the series}
Evaluating the sum (\ref{sumtoprove}) in the limit $s\to \infty$ proceeds in two steps. First, the asymptotic form of the terms 
\be \label{definnewD}
\Delta^s_{l,n} = \left(\frac{2l+1}{2s+1}\right)^{1/2} 
                 \clebsch{s}{s}{l}{s-n}{n-s}{0} 
\ee
to be summed is determined with the help of a recurrence formula for Clebsch-Gordan coefficients. Then, the sums are transformed into integrals which can be evaluated. All approximations drop terms of the order $1/s$ at least, hence the result is exact in the limit of infinite $s$.

Clebsch-Gordan coefficients satisfy the following recursion relation \cite{vilenkin69}:
\bea \label{CGrec}
&   & [ l(l+1) -2s(s+1) + 2m^2 ] \clebsch{s}{s}{l}{m}{-m}{0} \nonumber \\ 
&=& [s(s+1) - m(m+1)] \clebsch{s}{s}{l}{m+1}{-(m+1)}{0} \nonumber \\
&   &  + [s(s+1) - m(m-1)] \clebsch{s}{s}{l}{m-1}{-(m-1)}{0}  \, ,
\eea
implying that  
\bea \label{recforD}
(n+1) \left( 1 - \frac{n+1}{2s+1} \right) \Delta^s_{l,n+1} 
+ \left( 2n+1-\frac{2n^2+ 2n +1}{2s+1} \right) \Delta^s_{l,n}
+ & & \nonumber \\ 
\hspace{1cm} + n \left( 1 - \frac{n}{2s+1} \right) \Delta^s_{l,n-1} 
 = \frac{l(l+1)}{2s+1}  \Delta^s_{l,n} \, .
\eea
For any finite $n$ the terms subtracted on the left-hand-side 
become less and less important if $s\to \infty$. Assume now that one can 
write the terms with large values of $n$ in the form
\be \label{assumption}
\Delta^s_{l,n} (x_l) = \Lambda_n (x_l) \Delta^s_{l,0}  \, , \qquad 
\Lambda_0 (x_l) =1\, , \qquad 
 x_l = \frac{l(l+1)}{2s+1} \, .
\ee
The polynomial $\Lambda_n (x_l) $ of order $n$ in $x_l$ satisfies a three-term recursion relation,
\be \label{Lrecursion}
(n+1) \Lambda_{n+1} (x_l)  + (2n+1) \Lambda_{n} (x_l) + n \Lambda_{n-1} (x_l)
= x_l \Lambda_{n} (x_l) \, ,
\ee
where terms of order $1/s$ have been dropped in (\ref{Lrecursion}). Its solutions \cite{abramowitz+84} are proportional to the Laguerre polynomials, 
and the `normalization' condition $\Lambda_0 (x_l) =1$ implies that   
\be \label{laguerre}
\Lambda_{n} (x_l) 
     = (-)^n L_n (x_l) 
     = (-)^n\sum_{k=0}^n {n\choose k} \frac{(-x_l)^k}{k!} \, , \qquad 
n= 0,1,2, \ldots
\ee
The term $\Delta^s_{l,0}$ in (\ref{assumption}) can be determined in the following way. If $s$ is large, one writes for each finite $k$ 
\be \label{lowestterm}
\left( 1 - \frac{k}{2s+1}\right)^{2s+1}  \sim \exp [ -k ] \, , 
\ee
which leads to the approximation
\bea \label{lowest delta}
\Delta^s_{l,0} 
&=&  \left(\frac{2l+1}{2s+1}\right)^{1/2} \clebsch{s}{s}{l}{s}{-s}{0} 
  =  \frac{2l+1}{2s+1} \left( \frac{(2s)!}{(2s-l)!} \frac{(2s)!}{(2s+l+1)!}\right)^{1/2}
\nonumber \\
&=&  \frac{2l+1}{2s+1}  
     \left( \frac{\Pi_{k=0}^l (1- k/(2s+1) )}{\Pi_{k=0}^l (1+ k/(2s+1) )} \right)^{1/2} 
\sim \frac{2l+1}{2s+1} \exp \left[- \frac{1}{2} \frac{l(l+1)}{2s+1} \right] \, .
\eea
Collecting the results, one has 
\be \label{approxsum}
\lim_{s\to \infty} \, \, \sum_{l=0}^{2s} \Delta^s_{l,n} 
\sim  (-)^n \lim_{s\to \infty}  \sum_{l=0}^{2s} \Delta x_l  L_n (x_l)  e^{- x_l/2}  \, ,
\ee
where $\Delta x_l = (x_{l+1} - x_l) = (2l+1)/(2s+1)+ {\cal O} (1/s)$. Transforming now the Riemann sum into an integral, one obtains the final result  
\be \label{just2}
\lim_{s\to \infty} \, \, \sum_{l=0}^{2s} \Delta^s_{l,n} 
= (-)^n \int_{0}^{\infty} dx \,   L_n (x)  e^{- x/2} = 2 \, ,
\ee
using the formula %
\be \label{explag}
\int_{0}^{\infty} dx \,   L_n (x)  e^{- x/t} =  t (1-t)^n \, ,
\ee
for $t=2$. This identity is proven easily by means of the expansion in (\ref{laguerre}). 
\section{Discussion}
Starting from a new form of the kernel defining the familiar Wigner
formalism for a spin, its limit for infinite values of $s$ has been 
shown to be the Wigner kernel of a particle. As the kernel defines 
entirely a phase-space representation, this result guarantees that the 
Moyal formalism for a particle is reproduced automatically and {\em in toto},
if the limit $s\to \infty$ of the spin Moyal formalism is taken. 

In fact, slightly more has been shown. The result removes an ambiguity of the Moyal formalism for a spin: the Stratonovich-Weyl postulates are compatible with a discrete family of $2^{2s}$ distinct kernels $\widehat \Delta_\varepsilon ({\bf n})$. However, only {\em one} of these kernels turns into the particle kernel. This kernel had been singled out before for other reasons \cite{amiet+91}. In summary, the group theoretical contraction shows that the phase-space representations \`a la Wigner for spin and particle
systems are structurally equivalent.
\subsection*{Acknowledgements}
St. W. acknowledges financial support by the {\em Schweizerische
Nationalfonds}.
%

\end{document}